\documentclass[twocolumn,showpacs,prd]{revtex4}
\usepackage{mathrsfs,bm,multirow}
\usepackage{longtable,lscape}
\usepackage{txfonts}
\usepackage{amssymb}
\usepackage{indentfirst}
\usepackage{graphicx,,booktabs}
\usepackage{color}
\usepackage{amssymb}

\usepackage[dvipdfm,  
            ]{hyperref}

\begin{document}

\title{Does the enhancement observed in $\gamma\gamma\to D\bar{D}$ contain two $P$-wave higher charmonia?}
\author{Dian-Yong Chen$^{1,3}$}\email{chendy@impcas.ac.cn}
\author{Jun He$^{1,3}$}\email{junhe@impcas.ac.cn}
\author{Xiang Liu$^{1,2}$\footnote{corresponding author}}\email{xiangliu@lzu.edu.cn}
\author{Takayuki Matsuki$^4$}\email{matsuki@tokyo-kasei.ac.jp}
\affiliation{
$^1$Research Center for Hadron and CSR Physics,
Lanzhou University and Institute of Modern Physics of CAS, Lanzhou 730000, China\\
$^2$School of Physical Science and Technology, Lanzhou University, Lanzhou 730000,  China\\
$^3$Nuclear Theory Group, Institute of Modern Physics of CAS, Lanzhou 730000, China\\
$^4$Tokyo Kasei University, 1-18-1 Kaga, Itabashi, Tokyo 173-8602, Japan}
\date{\today}

\begin{abstract}

Solved is a new puzzle raised by the observation of an enhancement
structure $Z(3930)$ in $\gamma\gamma\to D\bar{D}$.
If categorizing
$Z(3930)$ as $\chi_{c2}(2P)$ suggested by Belle
and BaBar,
we must explain why $\chi_{c0}(2P)$ dominantly decaying into
$D\bar{D}$ is missing in the $D\bar{D}$ invariant mass spectrum. In
this work, we propose that the $Z(3930)$ enhancement structure may
contain two $P$-wave higher charmonia {$\chi_{c0}(2P)$}
and $\chi_{c2}(2P)$. We show that this assumption is supported by
our analysis of the $D\bar{D}$ invariant mass spectrum and
$\cos\theta^\ast$ distribution of $\gamma\gamma\to D\bar{D}$.
This observation would not only provide valuable information of two $P$-wave higher charmonia $\chi_{c0}(2P)$ and $\chi_{c2}(2P)$, but also serve as the crucial test of our novel proposal to the observed enhancement structure $Z(3930)$, especially at the forthcoming BelleII and the approved SuperB.
\end{abstract}


\maketitle

{\it Introduction: }
Since the Belle Collaboration reported the first observation of $X(3872)$ \cite{Choi:2003ue}, more and more charmonium-like states $XYZ$ have been observed in experiments \cite{charmoniu-like}. Both theoretical and experimental studies of charmonium-like states have become an intriguing and challenging research field of hadron physics.
It is well known that the potential model of hadron has made success for describing the low-lying hadron spectra \cite{Godfrey:1985xj} as well as heavy quarkonia. Meanwile, the abundant higher charmonium and bottomonium spectra were predicted. We believe these newly observed charmonium-like states have a close relationship with higher charmonia, which provides a good platform to perform the study of higher charmonium. It is valuable to reveal not only the underlying structures of these charmonium-like states, but also our deep understanding of non-perturbative behavior of QCD.

As listed in Particle data group (PDG) \cite{Nakamura:2010zzi}, there are three established $P$-wave spin-triplet charmonia, $\chi_{c0}(3415)$, $\chi_{c1}(3510)$, and $\chi_{c2}(3556)$. However, compared with the $S$-wave, the $P$-wave charmonium spectrum has not yet been established. The observations of charmonium-like states $X(3812)$ \cite{Choi:2003ue}, $Z(3930)$ \cite{Uehara:2005qd,Aubert:2010ab} and $X(3915)$ \cite{Uehara:2009tx} also have stimulated the theorists' extensive interest \cite{Kalashnikova:2005ui,Liu:2009fe,Danilkin:2010cc} in exploring the higher excitations of
these $P$-wave spin-triplet charmonia.

Together with these observations, we need to know details of $Z(3930)$ \cite{Uehara:2005qd,Aubert:2010ab}, which is a charmonium-like state observed by Belle \cite{Uehara:2005qd} and confirmed by BaBar \cite{Aubert:2010ab} in the $\gamma\gamma\to D\bar{D}$ process. Belle gave its mass $M=3929\pm5(\mathrm{stat})\pm2(\mathrm{syst})$ MeV with $\Gamma=29\pm10(\mathrm{stat})\pm2(\mathrm{syst})$ MeV \cite{Uehara:2005qd}, and BaBar obtained the consistent results with Belle, $M=3926.7\pm2.7(\mathrm{stat})\pm1.1(\mathrm{syst})$ MeV with $\Gamma=21.3\pm6.8(\mathrm{stat})\pm3.6(\mathrm{syst})$ MeV \cite{Aubert:2010ab}. Both Belle and BaBar indicated that $Z(3930)$ is the candidate of $\chi_{c2}(2P)$ suggested by their analysis of the angular distribution \cite{Uehara:2005qd,Aubert:2010ab}. Thus, the $Z(3930)$ experimentally established makes the $P$-wave spin-triplet charmonium spectrum abundant. In addition, it is not difficult to explain $X(3872)$ as the first radial excitation of $\chi_{c1}(3510)$ when considering the coupled channel effects \cite{Kalashnikova:2005ui,Danilkin:2010cc} or the dominant $2P$ $c\bar{c}$ component mixing with $D^0\bar{D}^{*0}+D^{*0}\bar{D}^0$ \cite{Meng:2005er}.

As the $P$-wave spin-triplet charmonium spectrum becomes
more abundant, however, an urgent and crucial question
emerges out of the study on the first radial excitation of $P$-wave
charmonia. {Very recently, the BaBar Collaboration \cite{Lees:2012xs}
confirmed the observation of $X(3915)$ in the $\gamma\gamma\to J/\psi\omega$ process and indicated that
$X(3915)$ is a $\chi_{c0}(2P)$ charmonium by a spin-parity analysis. This new observation is consistent with the prediction of
the property of $X(3915)$ given in Ref. \cite{Liu:2009fe}. Thus, these experimental measurements show that the mass of $\chi_{c0}(2P)[X(3915)]$ as
the first radial excitation of $\chi_{c0}(3415)$ is very close to
that of $\chi_{c2}(2P)[Z(3930)]$ and
above the $D\bar{D}$ threshold. Additionally, the decay behavior of $Z(3930)$ and $\chi_{c0}(2P)$ shows that both of them decay into
$D\bar{D}$ via the D-wave and S-wave interactions, respectively, where $\chi_{c0}(2P)\to D\bar{D}$ is a dominant
contribution to the total width \cite{Liu:2009fe}. Since $Z(3930)$
was already observed in the $D\bar{D}$ invariant mass spectrum of
the $\gamma\gamma\to D\bar{D}$ process
\cite{Uehara:2005qd,Aubert:2010ab}, we believe that $\chi_{c0}(2P)$ should
exist in the same data samples of the $D\bar{D}$ invariant mass
spectrum, where $\chi_{c0}(2P)$ and $Z(3930)$ are assumed to have the
same spatial wave functions.} However, the present experiment did not report any
evidence of $\chi_{c0}(2P)$ in this process
\cite{Uehara:2005qd,Aubert:2010ab}, which obviously contradicts the
above {general analysis}. This is a new puzzle when
studying the $P$-wave higher charmonia.

Due to the peculiarity of $\chi_{c0}(2P)$, in this work we propose a novel conjecture to solve this puzzle, i.e., the enhancement structure observed in the $D\bar{D}$ invariant mass spectrum of the $\gamma\gamma\to D\bar{D}$ process \cite{Uehara:2005qd,Aubert:2010ab} should contain both $\chi_{c0}(2P)$ and $\chi_{c2}(2P)$ signals. To testify that this conjecture is reasonable, in the following analysis we shall construct a model depicting the $\gamma \gamma \to D\bar{D}$ process, where we will consider both $\chi_{c0}(2P)$ and $\chi_{c2}(2P)$ contributions to this process. Furthermore, by fitting the experimental data of the $D\bar{D}$ invariant mass spectrum \cite{Uehara:2005qd,Aubert:2010ab}, we will test this conjecture and extract the resonance parameters of $\chi_{c0}(2P)$ and $\chi_{c2}(2P)$, which are important properties of $P$-wave higher charmonia. What is more important is that we find a complete series of $P$-wave spin-triplet charmonia including the established ground states and their corresponding first radial excitations.
Of course, this will inspire experimentalists' interest in studying these $\chi_{c0}(2P)$ and $\chi_{c2}(2P)$ in future experiment, too.

{\it Formulation: }
In general, $\gamma(k_1) \gamma(k_2) \to D(p_1)\bar{D}(p_2)$ occurs via two different mechanisms. The first mechanism is the so-called direct process, which provides the background contribution when studying the $D\bar{D}$ invariant mass spectrum. Its amplitude can be parameterized as
\begin{eqnarray}
\mathcal{A}\textnormal{\Huge{$[$}}
\raisebox{-15pt}{\includegraphics[width=0.12%
\textwidth]{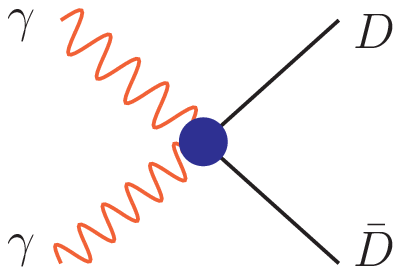}}\textnormal{\Huge{$]$}}_{NOR}= { g_{NoR} }\epsilon_1^\mu \epsilon_2^\nu \left( T^{0}_{\mu \nu} +
c_{02} T^2_{\mu \nu} \right ) \mathcal{F}(s),\label{dir}
\end{eqnarray}
which is abbreviated as $\mathcal{A}_{NOR}$. Here $T^0_{\mu \nu}= \mathcal{T}_{\mu \nu}$ and $T^2_{\mu \nu} =
\mathcal{E}_{\mu \nu, \alpha \beta} \mathcal{P}^{\alpha \beta \rho\lambda} p_{1\rho} p_{2\lambda}$ correspond to the helicity-0 and helicity-2 contributions, respectively, which can be distinguished by the superscripts $0$ and $2$.
 Here $\epsilon_1^\mu(k_1)$ and $\epsilon_2^\nu(k_2)$ are polarization vectors of two photons and the definitions of $\mathcal{T}_{\mu \nu}$, $\mathcal{E}_{\mu \nu, \alpha \beta}$, and $\mathcal{P}^{\alpha \beta \rho \lambda}$ will be given later.
 $\mathcal{F}(s)$ is the function of
the energy in the $\gamma\gamma$ center-of-mass frame, and is defined as
\cite{Aubert:2010ab}
\begin{eqnarray}
\mathcal{F}^2(s)=\sqrt{s-m_t^2} (\sqrt{s}-m_t)^\alpha \text{exp}
[-\beta (\sqrt{s}-m_t)]{/m_t^{\alpha+1} },
\end{eqnarray}
where $m_t$ is taken as the $D\bar{D}$ threshold. Free parameters
$\alpha$ and $\beta$ are determined by fitting the experimental
data. {In order to make the form factor dimensionless, a
factor $m_t^{-(\alpha+1)}$ is multiplied}.

The second mechanism is the intermediate resonance state contribution to the $\gamma \gamma \to D\bar{D}$ process, which is from the s-channel. The initial $\gamma \gamma$ and final $D\bar{D}$ are connected by the intermediate resonances with $J^P=0^+,2^+$, where the parity and total quantum number of the intermediate state are constrained by both the initial $\gamma\gamma$ and final $D\bar{D}$. When analyzing the data of Belle \cite{Uehara:2005qd} and BaBar \cite{Aubert:2010ab}, we select $\chi_{c0}(2P)$ and $\chi_{c2}(2P)$ as the intermediate states, which are the first radial excitations of $\chi_{c0}(3415)$ and $\chi_{c2}(3556)$, respectively.
These ground states can also contribute to $\gamma\gamma\to D\bar{D}$ as the intermediate resonances. However, since the masses of $\chi_{c0}(3415)$ and $\chi_{c2}(3556)$ are far away from the energy range discussed here, the high-mass tail of these resonances only provide one of the backgrounds to the $D\bar{D}$ invariant mass spectrum in the energy range ($>3.7$ GeV), which can be included in Eq. (\ref{dir}).

To investigate the process $\gamma\gamma\to D\bar{D}$ by the
intermediate $\chi_{c0}(2P)$ and $\chi_{c2}(2P)$, we adopt the
effective Lagrangian approach. Here, the interactions of
$\gamma\gamma$ and $D\bar{D}$ with {$\chi_{cJ}(2P)$} are
described by the effective Lagrangians
\begin{eqnarray}
\mathcal{L}_{\chi_{c0}(2P) \gamma \gamma} &=& g_{\chi_{c0}(2P)
\gamma \gamma} A^\mu A^\nu \mathcal{T}_{\mu \nu} \chi_{c0}(2P), \\
\mathcal{L}_{\chi_{c2}(2P) \gamma \gamma} &=& g_{\chi_{c2}(2P)
\gamma \gamma} A^\mu A^\nu \chi_{c2}^{\alpha \beta} (2P)
\mathcal{E}_{\mu \nu, \alpha \beta}, \\
\mathcal{L}_{\chi_{c0}(2P) D \bar{D}} &=& g_{\chi_{c0}(2P) D
\bar{D}} \chi_{c0}(2P) D\bar{D},\\
\mathcal{L}_{\chi_{c2}(2P) D \bar{D}} &=& -g_{\chi_{c2}(2P) D
\bar{D}} \chi_{c2}^{\mu \nu} (2P) \partial_{\mu} D \partial_{\nu}
\bar{D},
\end{eqnarray}
where
\begin{eqnarray}
\mathcal{T}^{\mu \nu} &=& g^{\mu \nu} -k_1^{\nu} k_{2}^{\mu}/ k_1 \cdot k_2,\\
\mathcal{E}^{\mu \nu, \alpha \beta} &=& g^{\mu \alpha}
g^{\nu \beta}-\frac{\left[ g^{\nu \beta} k_1^{\alpha} k_2^{\mu} + g^{\mu
\alpha} k_1^{\nu} k_2^{\beta}- g^{\mu \nu} k_1^{\alpha} k_2^{\beta}
\right]}{k_1 \cdot k_2},
\end{eqnarray}
in the momentum space.

The propagators of $\chi_{cJ}(2P)\, (J=0,2)$ are in the form,
 \begin{eqnarray*}
&&\chi_{c0}(2P):
\frac{i}{q^2-m_{\chi_{c0}(2P)}^2+im_{\chi_{c0}(2P)}
\Gamma_{\chi_{c0}(2P)}},\\
&&\chi_{c2}(2P):  \frac{i\, \mathcal{P}^{\mu\nu\mu^\prime
\nu^\prime}}{q^2-m_{\chi_{c2}(2P)}^2+im_{\chi_{c2}(2P)}\Gamma_{
\chi_{c0}(2P)}}
\end{eqnarray*}
with $\mathcal{P}_{\mu\nu\mu^\prime \nu^\prime}= \frac{1}{2}
(\tilde{g}_{\mu\mu^{\prime}} \tilde{g}_{\nu\nu^{\prime}}
+\tilde{g}_{\mu\nu^{\prime}} \tilde{g}_{\nu\mu^{\prime}})
-\frac{1}{3} \tilde{g}_{\mu\nu} \tilde{g}_{\mu^{\prime}
\nu^{\prime}}$ and  $\tilde{g}_{\mu \nu} = - g_{\mu \nu} +
{q_{\mu} q_{\nu} /q^2}$ .

In this mechanism, one obtains the amplitudes for $\gamma\gamma\to D\bar{D}$ via $\chi_{c0}(2P)$ and $\chi_{c2}(2P)$,
\begin{eqnarray}
\mathcal{A}\textnormal{\Huge{$[$}}
\raisebox{-15pt}{\includegraphics[width=0.20%
\textwidth]{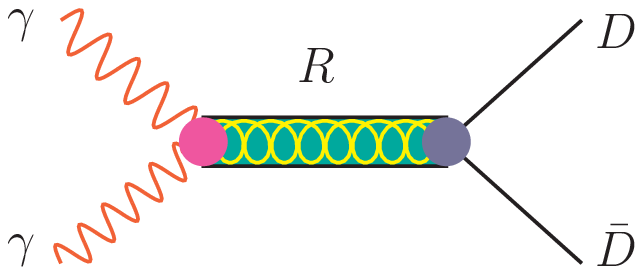}}
\textnormal{\Huge{$]$}}_R\equiv \mathcal{A}_R
\end{eqnarray}
with
\begin{eqnarray*}
\mathcal{A}_{\chi_{c0}(2P)} &=&
\frac{ig_{\chi_{c0}(2P) \gamma
\gamma} \epsilon_{1}^{\mu} \epsilon_{2}^{\nu} \mathcal{T}_{\mu \nu}g_{\chi_{c0}(2P)
D\bar{D}}}{s -m_{\chi_{c0}(2P)}^2+ i m_{\chi_{c0}(2P)}
\Gamma_{\chi_{c0}(2P)} } ,\\
\mathcal{A}_{\chi_{c2}(2P)} &=&  \frac{i g_{\chi_{c2}(2P) \gamma
\gamma} \epsilon_{1\mu} \epsilon_{2\nu} \mathcal{E}^{\mu \nu,
\alpha \beta} \mathcal{P}_{\alpha \beta \rho \lambda}}{s
-m_{\chi_{c2}(2P)}^2 + i m_{\chi_{c2}(2P)}
\Gamma_{\chi_{c2}(2P)} } \left(- g_{\chi_{c2}(2P) D\bar{D}}
ip_{1}^\rho ip_{2}^{\lambda}\right).
\end{eqnarray*}

The total amplitude of $\gamma\gamma\to D\bar{D}$ is written as
\begin{eqnarray}
\mathcal{M}_{Total} =\mathcal{A}_{NOR} +e^{i
\phi_0}\mathcal{A}_{\chi_{c0}(2P)} +e^{i\phi_2}
\mathcal{A}_{\chi_{c2}(2P)},\label{tot}
\label{decay}
\end{eqnarray}
where the background and intermediate $\chi_{c0}(2P)/\chi_{c2}(2P)$
resonance contributions are considered. Phases $\phi_0$ and $\phi_2$
are introduced to describe the the interference of the amplitudes.
Finally, one can compare the measured distributions with
\begin{eqnarray}
{\frac{dN}{dt}=N_0 \frac{d\sigma}{dt} = N_0
\frac{1}{64\pi s} \frac{1}{|\vec p_{1\mathrm{cm}}|^2}
\left|\mathcal{M}_{Total}\right|^2,\label{total} }
\end{eqnarray}
where $s=q^2\equiv(k_1+k_2)^2$ is the square of the energy, $\vec
p_{1\mathrm{cm}}$ is the momentum of $D$ in the center-of-mass frame
and $N_0$ a normalization factor. In addition, we define
$$t=(k_1-p_1)^2= m_D^2-\frac{s}{2}+\frac{1}{2}\sqrt{s(s-4m_D^2)}\cos \theta^\ast.$$ And $\theta^\ast$ denotes the angle between outgoing $D$ meson
and incoming photon in the center-of-mass frame.

{\it Numerical Calculation : }
As presented in Refs. \cite{Uehara:2005qd,Aubert:2010ab}, the experimental data of the $D\bar{D}$ invariant mass spectrum and $\cos\theta^\ast$ distribution of $\gamma\gamma\to D\bar{D}$ were released. Based on these data, with the help of Eq. (\ref{total}) we study whether this enhancement structure observed by Belle and BaBar contains both $\chi_{c0}(2P)$ and $\chi_{c2}(2P)$ resonances.

\begin{table}[htb]
\centering %
\caption{The values of these parameters obtained by fitting the
Belle data \cite{Uehara:2005qd}. {The resonance
parameters masses and decay widths of $\chi_{c0}(2P)$ and
$\chi_{c2}(2P)$ are in units of GeV and MeV, respectively. The phase
angles, $\alpha$ and $f_{\chi_{c2}(2P)}$ are dimensionless. $\beta$
and  $f_{\chi_{c0}(2P)}$ are in units of GeV$^{-1}$ and GeV$^{2 }$},
respectively. \label{Tab.Para}}
\begin{tabular}{cccc}
\hline\hline%
Parameters&Values&Parameters&Values\\\midrule[1pt]
$m_{\chi_{c0}(2P)} $       & $3.920    \pm       0.007$       &
$\Gamma_{\chi_{c0}(2P)} $  & $8.065    \pm       9.663$      \\
$m_{\chi_{c2}(2P)} $       & $3.942    \pm       0.003$      &
$\Gamma_{\chi_{c2}(2P)} $  & $11.980   \pm       6.953$     \\
$f_{\chi_{c0}(2P)} $       & $0.035    \pm       0.023$      &
$\phi_{0} $  (Rad)   & $2.919    \pm       0.287$      \\
$f_{\chi_{c2}(2P)} $       & $2.836    \pm       1.252$      &
$\phi_{2}  $ (Rad)   & $2.492    \pm       0.390$      \\
$\alpha $                     & $-0.497   \pm       0.788$      &
$\beta  $                     & $8.996    \pm       1.172$      \\
$c_{02}$                      & $39.996   \pm      11.509$      &
\\
\hline\hline%
\end{tabular}
\end{table}

In our model, {the lineshape of the $D\bar{D}$ invariant
mass spectrum and the $\cos\theta^\ast$ distribution are determined
by the following 11 parameters:}
\begin{eqnarray*}
&&m_{\chi_{c0}(2P)},\,\Gamma_{\chi_{c0}(2P)},\,m_{\chi_{c2}(2P)},\,
\Gamma_{\chi_{c2}(2P)},f_{\chi_{c0}(2P)},\,f_{\chi_{c2}(2P)},\\
&&\alpha,\,\beta,\,c_{02},\,\phi_0,\,\phi_2,
\end{eqnarray*}
{where $f_{\chi_{c0}(2P)}=
g_{\chi_{c0}(2P)\gamma\gamma}g_{\chi_{c0}(2P) D\bar{D}}/g_{NoR}$ and
$f_{\chi_{c2}(2P)}=g_{\chi_{c2}(2P)\gamma\gamma}g_{\chi_{c2}(2P)
D\bar{D}} /g_{NoR} $ and the coupling constants $g_{NoR}$ can be
absorbed by the normalization factor $N_0$. By fitting the
experimental data of the $D\bar{D}$ invariant mass spectrum and
$\cos\theta^\ast$ distribution of $\gamma\gamma\to D\bar{D}$
\cite{Uehara:2005qd,Aubert:2010ab}, these parameters can be
determined. Among all fitting parameters, $m_{\chi_{cJ}(2P)}$ and
$\Gamma_{\chi_{cJ}(2P)}$ ($J=0,2$) are important resonance
parameters reflecting the properties of $\chi_{c0}(2P)$ and
$\chi_{c2}(2P)$.}

With the above preparation, we perform a fitting of our model to the
Belle's data \cite{Uehara:2005qd}. Here, the reason why we choose
the Belle's data is that its information is more abundant than that
from BaBar \cite{Uehara:2005qd,Aubert:2010ab}. For instance, Belle
gave the $\cos\theta^\ast$ distribution in the
$3.91<M(D\bar{D})<3.95$ GeV region of $\gamma\gamma\to D\bar{D}$
including only the background contribution (the cyan histogram in
Fig. \ref{Fig:theta}) \cite{Uehara:2005qd}, which enables us to
constrain $c_{02}$ in Eq. (\ref{dir}) appropriately. {The $\cos\theta^\ast$ distribution of the background extracted by
the Belle collaboration depends on $\cos \theta^\ast$, which
indicates that there is helicity-2 component in the background and
this component will interfere with the resonances contributions from
$\chi_{c2}(2P)$ state and such interference is not included in the
experimental analysis \cite{Uehara:2005qd}.}

\begin{figure}[h!]
\centering \scalebox{0.66}{\includegraphics{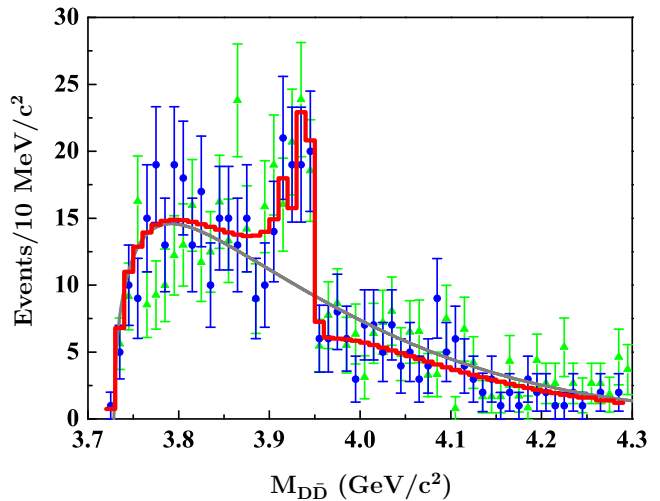}}
\caption{(color online). The best fit (red histogram) to the experimental data of the $D\bar{D}$
invariant mass distributions given by Belle \cite{Uehara:2005qd} (blue dots with
error bar) and BaBar \cite{Aubert:2010ab} (green triangles with error bar). Here, the
grey solid curve is the description of the background of $\gamma \gamma \to
D\bar{D}$ with these obtained parameters listed in Table \ref{Tab.Para} and Eq. (\ref{dir}). }\label{Fig:mDD}
\end{figure}

In Table \ref{Tab.Para}, we have listed the values of  parameters corresponding to our best fit to the Belle's data. Fig. \ref{Fig:mDD} shows the comparison of theoretical line shape of the $D\bar{D}$ invariant mass spectrum of $\gamma\gamma\to D\bar{D}$ with the experimental data. Although we fit our model to the Belle's data in the analysis of the $D\bar{D}$ invariant mass spectrum, we also make a comparison of our results with the BaBar's data. We need to normalize the $m_{D\bar{D}}$ invariant mass
spectrum distribution given by BaBar to the Belle's data by
multiplying the factor, which is the ratio of the total events measured by
Belle to those obtained by BaBar for the corresponding bins. As shown in Fig. \ref{Fig:mDD}, the obtained line shape of the $D\bar{D}$ invariant mass spectrum of $\gamma\gamma\to D\bar{D}$ can indeed describe these experimental data. In Fig. \ref{Fig:mDD}, we also show the line shape of the background (grey solid curve) when keeping only $\mathcal{A}_{NOR}$ contribution and adopting the optimum values of $\alpha$, $\beta$, and $c_{02}$ listed in Table \ref{Tab.Para} as the input.

We need to emphasize that a very steep line shape at $M_{D\bar{D}}\sim 3.95$ GeV exists in the experimental data of Belle \cite{Uehara:2005qd} and BaBar \cite{Aubert:2010ab}, which is reflected in our fit. When including $\chi_{c0}(2P)$ and $\chi_{c2}(2P)$, we can also reproduce the enhancement structure in the $D\bar{D}$ invariant mass spectrum of $\gamma\gamma\to D\bar{D}$ \cite{Uehara:2005qd,Aubert:2010ab}. Especially, the extracted resonance parameters of $\chi_{c0}(2P)$ and $\chi_{c2}(2P)$ indicate that these are two narrow resonances with masses very close to each other. These optimum resonance parameters are also consistent with the theoretical predictions of their masses \cite{Kalashnikova:2005ui,Danilkin:2010cc} and widths \cite{Liu:2009fe}.

\begin{figure}[h!]
\centering \scalebox{0.66}{\includegraphics{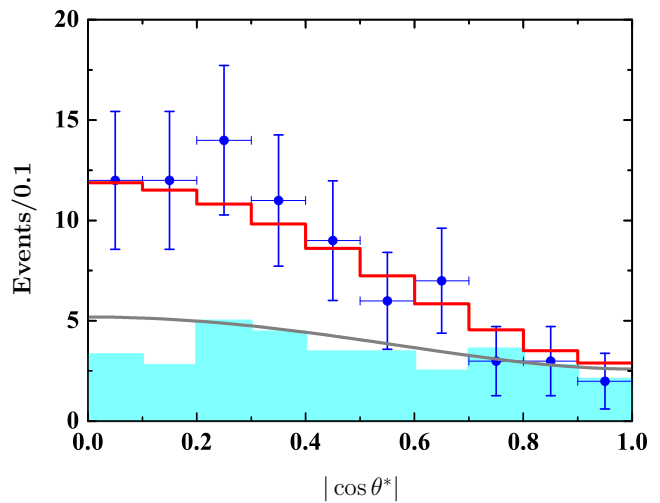}}
\caption{(color online).Our best fit (red histogram) to the $\cos \theta^\ast$
distribution of $\gamma\gamma\to D\bar{D}$. Here, we adopt the same parameters as those used in Fig. \ref{Fig:mDD}.
Blue dots with error bars are the Belle's data \cite{Uehara:2005qd}. The cyan histogram is the background measured by Belle \cite{Uehara:2005qd},
while the grey solid curve is our best fit to this
background. }\label{Fig:theta}
\end{figure}

Figure \ref{Fig:theta} shows our fit to the $\cos\theta^\ast$ distribution of $\gamma\gamma\to D\bar{D}$ given by Belle \cite{Uehara:2005qd}. Using our model with the parameters in Table \ref{Tab.Para}, we further get the $\cos\theta^\ast$ distribution in the $3.91<M(D\bar{D})<3.95$ GeV region of $\gamma\gamma\to D\bar{D}$ corresponding to Fig. \ref{Fig:mDD}. Comparing our results with the Belle's data, one finds that the data on $\cos\theta^\ast$ released by Belle can be well depicted even if the enhancement structure in the $D\bar{D}$ invariant mass spectrum is formed by  two $P$-wave higher charmonia $\chi_{c0}(2P)$ and $\chi_{c2}(2P)$.
In Refs. \cite{Uehara:2005qd,Aubert:2010ab}, both Belle and BaBar once claimed that the enhancement structure observed in the $D\bar{D}$ invariant mass spectrum is due to only one resonance with $J^{PC}=2^{++}$ as the candidate of $\chi_{c2}(2P)$ by performing the angular distribution analysis \cite{Uehara:2005qd,Aubert:2010ab},
while we claim that two resonances can explain the same enhancement structure as shown by detailed numerical calculations in this paper.

\begin{figure*}[htbp]
\centering \scalebox{0.9}{\includegraphics{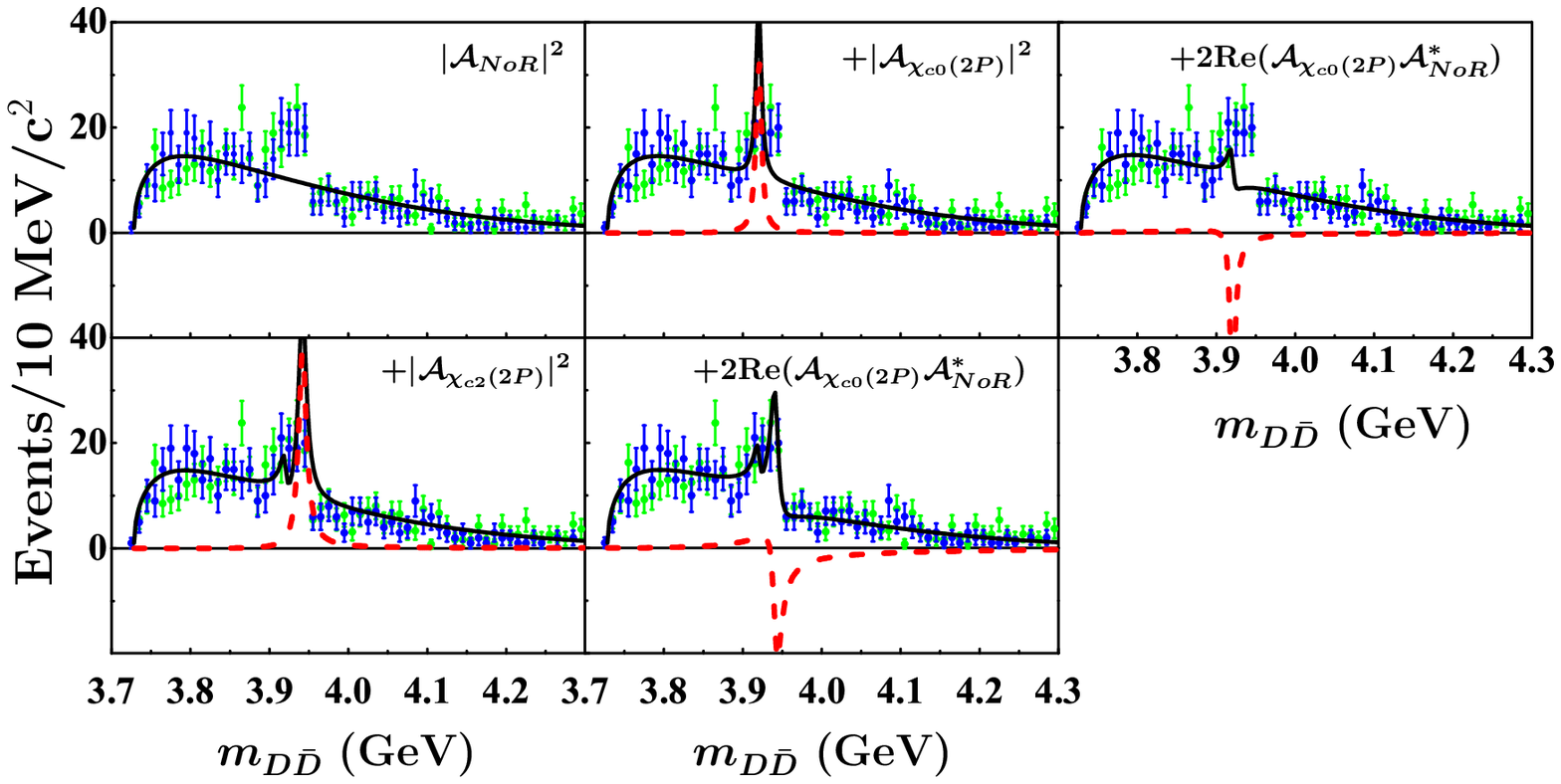}}\\
\centering \scalebox{0.9}{\includegraphics{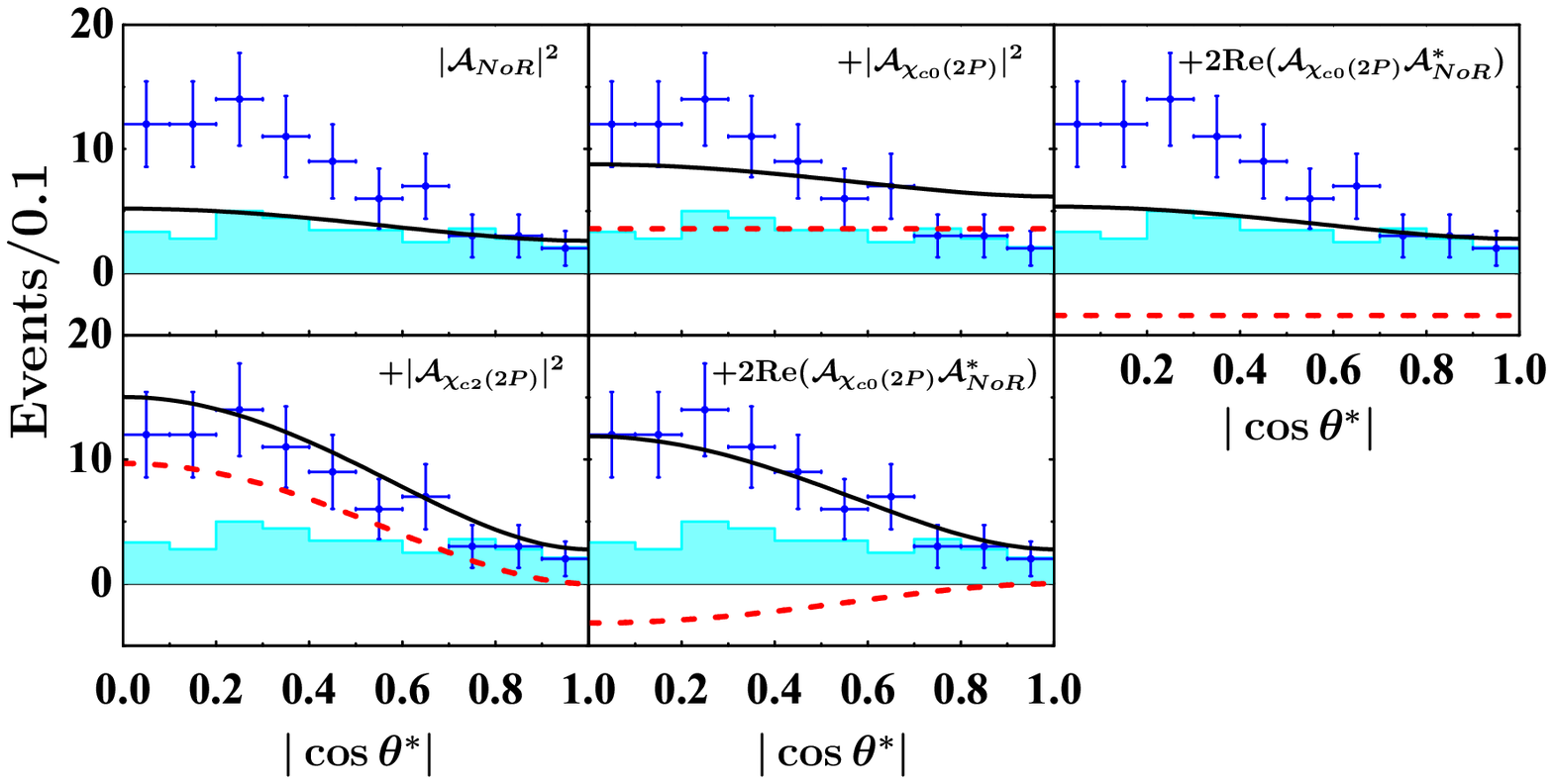}}
\caption{(color online). {The
distribution of the $D\bar{D}$ invariant mass spectrum and the
$\cos\theta^*$ distribution of $\gamma\gamma\to D\bar{D}$. Here, we
present the evolution of the theoretical result (black solid curves)
by adding the contributions from five terms:
($|\mathcal{A}_{NoR}|^2$, $|\mathcal{A}_{\chi_{c0}(2P)}|^2$,
$2\mathrm{Re}(\mathcal{A}_{\chi_{c0}(2P)}\mathcal{A}^*_{NoR})$,
$|\mathcal{A}_{\chi_{c2}(2P)}|^2$,
$2\mathrm{Re}(\mathcal{A}_{\chi_{c2}(2P)}\mathcal{A}^*_{NoR})$) one
by one. The red dashed curve in each diagram reflects the
contribution of each of five terms in $|\mathcal{M}_{Total}|^2$ in
Eq. (\ref{tot}).}}\label{Fig:step}
\end{figure*}

{Although $f_{\chi_{c0}(2P)}$ is far smaller than $f_{\chi_{c2}(2P)}$ as shown in Table \ref{Tab.Para}, we cannot conclude that the $\chi_{c0}(2P)$ contribution can be ignored because it should be emphasized that the dimension of $f_{\chi_{c0}(2P)}$ is different from that of $f_{\chi_{c2}(2P)}$. To further explain this point, in Fig. \ref{Fig:step} we present the evolution of the distribution of the $D\bar{D}$ invariant mass spectrum and the $\cos\theta^*$ distribution by adding the $\chi_{c0}(2P)$ and $\chi_{c2}(2P)$ contributions step by step, where we adopt the the central values of these parameters listed in Table \ref{Tab.Para}. In Fig. \ref{Fig:step}, we can find that $\chi_{c0}(2P)$ is indeed important to reproduce the experimental data. Thus, the small value of $f_{\chi_{c0}(2P)}$ does not mean that the $\chi_{c0}(2P)$ contribution is small and can be ignored.
{Here, we also give the ratio for the cross section of $\gamma\gamma \to D\bar{D}$ via $\chi_{c2}(2P)$ to that via $\chi_{c0}(2P)$, which is 1.447 due to our calculation. This ratio further shows that the $\chi_{c0}(2P)$ contribution cannot be ignored although the coupling $f_{\chi_{c0}(2P)}$ is small.} }

{\it Summary: }
An enhancement structure named as $Z(3930)$ in $\gamma\gamma\to D\bar{D}$ reported by Belle \cite{Uehara:2005qd} and confirmed by BaBar \cite{Aubert:2010ab} raises a new puzzle in the investigation of $P$-wave higher charmonium. If $Z(3930)$ is a good candidate of $\chi_{c2}(2P)$ charmonium as claimed by Belle and BaBar, we must explain why the signal of $\chi_{c0}(2P)$ is missing in the measured $D\bar{D}$ invariant mass spectrum of $\gamma\gamma\to D\bar{D}$ \cite{Uehara:2005qd,Aubert:2010ab} because the experimental study of $P$-wave charmonia indicated that the mass of $\chi_{c0}(2P)$ is close to that of $\chi_{c2}(2P)$ \cite{Lees:2012xs} and $\chi_{c0}(2P)$ dominantly decays into $D\bar{D}$ \cite{Liu:2009fe}. In this paper, we have proposed that the observed structure $Z(3930)$ should be formed by two $P$-wave higher charmonia $\chi_{c0}(2P)$ and $\chi_{c2}(2P)$. To examine whether this conjecture is reasonable, we have performed the analysis of
the $D\bar{D}$ invariant mass spectrum and $\cos\theta^\ast$ distribution of $\gamma\gamma\to D\bar{D}$. Our study has illustrated that the experimental data of these quantities can be well reproduced when considering both $\chi_{c0}(2P)$ and $\chi_{c2}(2P)$ in the $\gamma\gamma\to D\bar{D}$ process, which supports our conjecture of $Z(3930)$ structure. As indicated in Ref. \cite{babar-X3915}, BaBar is carrying out the spin-parity analysis of $X(3915)$, which will be helpful to clarify whether it is suitable to explain $X(3915)$ as $P$-wave higher charmonium $\chi_{c0}(2P)$.

Beside testing their resonance parameters by different phenomenological models,
experimental examination of whether there exist $\chi_{c0}(2P)$ and $\chi_{c2}(2P)$ in the $\gamma\gamma\to D\bar{D}$ data will be an important and interesting research topic in future experiments, especially in Belle, BaBar, forthcoming BelleII \cite{belle} and SuperB \cite{superb}. We expect further theoretical and experimental studies on these two $P$-wave higher charmonia.

Note added in proof : After completing our paper, there appears a
paper (arXiv:1207.2651),
in which the BaBar collaboration claims that they have confirmed the
existence of the charmonium-like resonance $X(3915)$ decaying to
$J/\psi \omega$ with a spin-parity assignment $J^P=0^+$
\cite{Lees:2012xs}, i.e., the identification of the signal as due to
the $\chi_{c0}(2P)$ that we have claimed in this paper as well as
in Ref. \cite{Liu:2009fe}.

\section*{Acknowledgment}
This project is supported by the National Natural Science Foundation of
China under Grants 11222547, 11175073, 11035006, 11005129, 10905077, the
Ministry of Education of China (FANEDD under Grant No. 200924,
DPFIHE under Grant No. 20090211120029, NCET, the Fundamental
Research Funds for the Central Universities), the Fok Ying-Tong Education Foundation (No. 131006) and the West Doctoral Project of Chinese Academy of Sciences.

\end{document}